\documentclass[final]{aipproc}

\layoutstyle{6x9}

\begin{document}

\title{Generalized parton distributions, the hunt for quark orbital momenta}


\author{E. Voutier}{
  address={Laboratoire de Physique Subatomique et de Cosmologie \\
  IN2P3/CNRS - Universit\'e Joseph Fourier - INP \\
  53 rue des Martyrs, 38026 Grenoble, France \\
  E-mail address: voutier@lpsc.in2p3.fr}
}

\begin{abstract}
The Generalized Parton Distributions (GPDs) are the appropriate framework for a 
universal description of the partonic structure of the nucleon. They characterize 
the dynamics of quarks and gluons inside the nucleon and consequently contain 
information about the spin of the nucleon. The current experimental knowledge about 
GPDs is rewieved with the emphasis on the determination of $E^q(Q^2,x,\xi,t)$, the 
least known and constrained GPD, of particular importance in the nucleon spin 
puzzle. The perspectives of this experimental program are also addressed. 
\end{abstract}

\maketitle


\section{INTRODUCTION}

After several decades of intensive experimental efforts, the origins of the nucleon 
spin still keep a lot of secrets. While the recent measurements at CERN, DESY, JLab 
and SLAC have clearly established the quark helicity contributions to the nucleon 
spin~\cite{Air07}, little is known about the gluon polarization 
contribution~\cite{Ell08}, and the quark and gluon orbital momenta contributions are 
essentially unknown. On the 
theoretical front, it is only recently that a comprehensive picture of the nucleon 
spin sum rules, enlightening the importance of the transversity distributions, was 
proposed~\cite{Bak04}. Intuitively, these different distributions are the expression 
of a single unique reality, the dynamics of quarks and gluons constituting the 
nucleon, which should rely on more fundamental quantities. Generalized Parton 
Distributions (GPDs)~\cite{{Mul94},{Rad97}} provide a universal and powerful 
framework which unifies in the same formalism electromagnetic form factors, parton
distributions, and the spin of the nucleon. GPDs encode the correlations between 
quarks, anti-quarks and gluons, and can be interpreted as the transverse 
distribution of partons carrying a certain longitudinal momentum fraction 
of the nucleon~\cite{{Bur00},{Die02}}, providing then a natural 
link with the transverse degrees of freedom. The GPD framework and its experimental 
knowledge is hereafter presented, particularly in the context of the nucleon 
spin puzzle.

\section{GENERALIZED PARTON DISTRIBUTIONS}

GPDs are universal non-perturbative objects entering the description of hard
scattering processes. They are defined for each quark flavor and gluon, 
and correspond to the amplitude for removing a parton of longitudinal 
momentum fraction $x+\xi$ and restoring it with momentum fraction $x-\xi$ 
(fig.~\ref{fig:hdbg}). In this process, the nucleon receives a four-momentum 
transfer $t=\Delta^2$ whose transverse component $\Delta_{\perp}$ is Fourier 
conjugate to the transverse position of partons. At the leading twist, the partonic 
structure of the nucleon~\cite{{Die03},{Bel05}} is described by four quark helicity conserving and chiral even GPDs ($H^q, 
\widetilde{H}^q, E^q,\widetilde{E}^q$) and four quark helicity flipping and chiral 
odd GPDs ($H^q_T, \widetilde{H}^q_T, E^q_T,\widetilde{E}^q_T$), together with eight 
similar gluon GPDs. In the forward limit ($t \to 0 , \xi \to 0$), the optical 
theorem links the $H$ GPDs to the usual density, helicity, and tranversity 
distributions measured in deep inelastic scattering (DIS). The $E$ GPDs, which 
involve a flip of the nucleon spin, do not have any DIS equivalent and consequently 
constitute a new piece of information about the nucleon structure. The first Mellin
moments relate chiral even GPDs to form factors, as $E^q$ with the Pauli
electromagnetic form factor 
\begin{equation}
\int_{-1}^{+1} dx \,\, E^q(Q^2,x,\xi,t) = F^q_2(t) \, ,
\end{equation}
and the second Mellin moments relate GPDs to the nucleon dynamics, particularly the
total angular momentum carried by the partons, following Ji's sum rule~\cite{Ji97}  
\begin{equation}
J^q = \frac{1}{2} \Delta \Sigma + L^q = \frac{1}{2} \int_{-1}^{+1} dx \,\, x \left[ 
H^q(Q^2,x,\xi,0) + E^q(Q^2,x,\xi,0) \right] \, .
\end{equation}
$\Delta \Sigma$ being known from DIS, GPD measurements allow to access the 
contribution of the orbital momemtum to the nucleon spin. In this prospect, 
$E(Q^2,x,\xi,t)$ is of particular interest since it is not constrained by any DIS
limit and remains essentially unknown. Similar relations have been proposed which 
relate chiral odd GPDs to the transverse spin-flavor dipole moment and the 
correlation between quark spin and angular momentum in an unpolarized 
nucleon~\cite{Bur05}.

\section{DEEPLY VIRTUAL EXCLUSIVE SCATTERING}

\begin{figure}[t]
\centering
\includegraphics[height=.16\textheight]{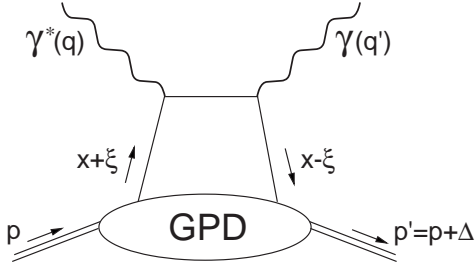}
\caption{Lowest order (QCD) amplitude for the virtual Compton process. The 
four-momentum of the incident and scattered photon are $q$ and $q'$. 
The four-momentum of the initial and final proton are $p$ and $p'$, with $\Delta$=$
(p'-p)$=$(q-q')$. The DIS scaling variable is $x_{\rm B}$=$Q^2/(2 p \cdot q)$ and 
the DVCS scaling variable is $\xi$=$x_{\rm B}/(2-x_{\rm B})$.}
\label{fig:hdbg}
\end{figure}

Deeply Virtual Compton Scattering (DVCS), corresponding to the absorption of a 
virtual photon by a quark followed quasi-instantaneously by the emission of a real 
photon, is the simplest reaction to access GPDs. In the Bjorken regime, $-t \ll Q^2$ 
and $Q^2$ much larger than the quark confinement scale, the leading contribution to 
the reaction amplitude is represented by the so-called handbag diagram 
(fig.~\ref{fig:hdbg}) which figures the convolution of a known $\gamma^{\ast} 
q \to \gamma q$ hard scattering kernel with an unknown soft matrix element 
describing the partonic structure of the nucleon parametrized by 
GPDs~\cite{{Ji98},{Col99}}. Consequently, GPDs ($E^q$) enter the reaction 
cross section through a Compton form factor ($\cal E$) which involves an 
integral over the intermediate quark propagators
\begin{eqnarray}
{\cal E} & = & \sum_q e_q^2 \, \, {\cal P} \int_{-1}^{+1} dx \, \left( \frac{1}{x-\xi} + 
\frac{1}{x+\xi} \right) E^q(Q^2,x,\xi,t) \label{eq:cff} \\
& \phantom{-} & \hspace*{70pt} -i \pi \sum_q e_q^2 \left[ E^q(Q^2,\xi,\xi,t) - 
E^q(Q^2,-\xi,\xi,t) \right] \nonumber
\end{eqnarray}
$e_q$ being the quark electric charge in units of the elementary charge. In addition 
to the DVCS amplitude, the cross section for electroproduction of photons gets 
contributions from the Bethe-Heitler (BH) process where the real photon is emitted 
by the initial or final lepton, leading to
\begin{equation}
\frac{d^5 \sigma}{dQ^2 dx_B dt d\phi_e d \varphi} = {\cal T}_{BH}^2 + {\vert 
{\cal T}_{DVCS} \vert}^2 \mp 2 \, {\cal T}_{BH} \Re e\{ {\cal T}_{DVCS} \}
\label{eq:sum}
\end{equation}
with $\phi_e$ the scattered electron azimuthal angle, and $\varphi$ the 
out-of-plane angle between the leptonic and hadronic planes; the $+(-)$ sign of the
interference amplitude stands for negative(positive) leptons. Though undistinguishable from DVCS, 
the BH amplitude is known and exactly calculable from the electromagnetic form 
factors. Beam charge and beam and/or target polarization observables can be 
advantageously used to select different contributions to the cross section. For 
instance, the polarized cross section difference for opposite beam helicities 
allows to isolate the imaginary part of the DVCS amplitude
\begin{eqnarray}
\frac{d^5 \Delta \sigma}{dQ^2 dx_B dt d\phi_e d \varphi} & = & \frac{1}{2} 
\left[ \frac{d^5 \overrightarrow{\sigma}}{dQ^2 dx_B dt d\phi_e d \varphi} - 
\frac{d^5 \overleftarrow{\sigma}}{dQ^2 dx_B dt d\phi_e d \varphi} \right] 
\label{eq:dif} \\
& = & {\cal T}_{BH} \, \Im m \{{\cal T}_{DVCS}\} + \Re e \{{\cal T}_{DVCS}\} \,
\Im m \{{\cal T}_{DVCS}\} \nonumber
\end{eqnarray}
where $\Im m \{{\cal T}_{DVCS}\}$ appears now linearly and magnified by the 
BH amplitude. In practice, the measurements of the cross section and a selected set 
of single and double spin polarization observables allows to extract the real an
imaginary parts of the Compton form factors, from which the GPDs can be 
deconvoluted~\cite{Gui08}. 

Deeply Virtual Meson Production (DVMP), where the real photon is replaced by a meson, 
is another channel to access GPDs which provides in addition an elegant flavor 
decomposition. In this case, the factorization of the cross section applies only to 
longitudinal virtual photons and the GPDs entering the Compton form factors 
(eq.~\ref{eq:cff}) are further convoluted with a meson distribution amplitude. The 
measurement of the angular distribution of the decay products of the vector mesons
allows to extract the longitudinal part of the cross section and the longitudinal 
polarization of the vector mesons, assuming the $s$-channel helicity conservation.
Other reaction mechanisms, like the 2-gluon exchange from a $q {\bar q}$ fluctuation 
of the virtual photon, may contribute to the production process. Similarly to DVCS,
polarization observables help to single-out the pure handbag contributions.

\section{EXPERIMENTAL STATUS}

The pioneering studies of the electro-production of photons at
DESY~\cite{{Air01},{Adl01}} and JLab~\cite{Ste01} did prove the existence of the 
DVCS mechanism by measuring sizeable beam spin asymmetries (BSA from the ratio of 
eq.~\ref{eq:dif} and eq.~\ref{eq:sum}) in the valence region and significant cross 
sections in the gluon sector. Other limited studies showed the importance of the 
beam charge~\cite{Air07-1} and the target polarization observables~\cite{Che06}.
The recent remarkable results of the starting dedicated DVCS experimental program of 
JLab is the strong indication for scaling in the valence region at $Q^2$ as low as 
$2$~GeV$^2$, and the observation of an unexpected DVCS amplitude magnitude at JLab
energies~\cite{Mun06}. This early scaling is also supported by the $\varphi$ angular 
dependence of the BSA measured at JLab with an unprecedented accuracy over a wide 
kinematic range~\cite{Gir08}. In general, GPD based calculations provide a
reasonable but unsatisfactory agreement with these data which turn out to be fairly
reproduced by a more conventional Regge approach~\cite{Lag07}. The significance of
this duality has not yet been resolved. 

In the meson sector, the experimental status with respect to GPDs remains 
controversial. Sizeable BSAs have been reported for exclusive neutral pion 
electro-production at JLab energies~\cite{DeM08}, which suggest that both longitudinal 
and transverse amplitudes contribute to the process. On the one hand, this forbids
a direct GPD based interpretation of the data, and on the other hand, a Regge based 
approach fails to reproduce them. The longitudinal cross section for the 
electro-production
of longitudinally polarized neutral rho was recently measured at JLab~\cite{Mor08}. 
Standard GPD calculations, particularly successful at high energies, do not reproduce
data in the valence region, while calculations based on hadronic degrees of freedom
are in very good agreement over the complete energy range scanned by the world data.
The observation that strongly modified GPDs allow for a partonic interpretation of 
these measurements raises the question whether current GPD parametrizations must be
revisited or the existence of strong higher twist corrections percludes a 
GPD wise interpretation of DVMP in the valence region.

\section{PERSPECTIVES}

\begin{figure}[t]
\centering
\includegraphics[height=.37\textheight]{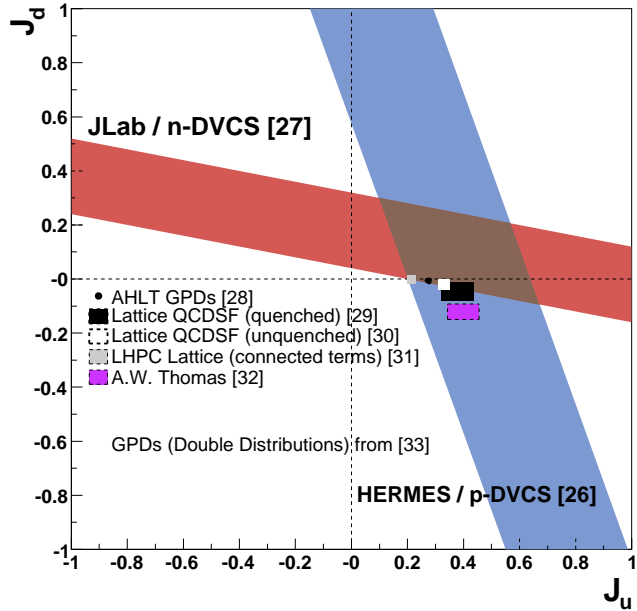}
\caption{Model dependent experimental constraint on $J_u$ and $J_d$ quark angular 
momenta from the HERMES $\vec{\mathrm{p}}$-DVCS~\cite{Air08} and the JLab 
n-DVCS~\cite{Maz07} experiments. Different model 
calculations~\cite{{Ahm07},{Goc04},{Sch07},{Hag07},{Tho08}} are compared to an 
extrapolation of experimental data within the VGG double distribution description
of GPDs~\cite{Van99}.}
\label{fig:jujd}
\end{figure}
Transversely polarized proton targets are of primary importance for the determination
of the quark orbital momenta, since they provide a direct access to the GPD $E^q$ 
which is essentially parametrized in terms of the quark angular momenta~\cite{Ell06}. 
In this respect, DVCS and DVMP transverse target asymmetries (TSA) are key 
observables. The first experimental indication of this sensitivity was reported by 
HERMES from DVCS TSA measurements off a proton target~\cite{Air08}. Within a simple 
twist-2 approach, the TSA data may be interpreted as a measurement of the GPDs 
combination $\approx t [F_2 H - F_1 E] / 4 M^2$. Relying on a given GPD model, the 
most probable parameters $J_u$ and $J_d$, representing the $u$ and $d$ quarks angular 
momenta, can be extracted from the comparison between data and calculations, leading 
to the vertical band in fig.~\ref{fig:jujd}. \\
Another important observable connected to $E^q$ is the DVCS polarized cross section 
difference (eq.~\ref{eq:dif}) off a neutron target. Taking advantage of the smalness
of the neutron Dirac electromagnetic form factor and of the cancellation between the 
$u$ and $d$ quarks in the GPD $\widetilde{H}$ at small $t$, this observable is, at 
leading twist, a measurement of the combination $t F_2 E / 4 M^2$. The first data 
on this observable were obtained by the JLab n-DVCS experiment~\cite{Maz07}.
Similarly to TSA, the comparison with calculations for the same GPD model leads to 
the horizontal band in fig.~\ref{fig:jujd}. \\
In the present status of GPD models, the main message of this appealing picture is 
the experimental demonstration of the complementarity between transversely polarized 
proton and neutron targets in the search for quark orbital momenta. This fundamental 
feature is a direct consequence of isospin symmetry. Experimental advances in this 
quest are depending on the progress in the developments of high luminosity 
transversely polarized targets~\cite{San08} and on the next generation of n-DVCS
experiments for which transversely polarized targets may provide an interesting 
observable to extract $E$ at very small $t$~\cite{Vou08}.

\section{CONCLUSIONS}

The exciting exploration of the GPDs world is progressing. A larger set of 
experimental data are expected from JLab 6 GeV and the HERMES device supplemented
by a recoil detector. In a near future, the energy upgrade of JLab and the completion
of a DVCS program at COMPASS will considerably enlarge the kinematic range of GPDs
knowledge. Together with the evolution of lattice QCD calculations capabilities, one
may reasonably expect to get a comprehensive and quantitative understanding of the
nucleon structure, including the contribution of the quark orbital momenta to the 
nucleon spin, by the end of the next decade.   

\begin{theacknowledgments}

I would like to thank F.-X.~Girod, M.~Guidal, N.~d'Hose, M.~Mazouz and A.~Sandorfi
for providing pictures and material for this presentation. This work was supported 
in part by the U.S. Department of Energy (DOE) contract DOE-AC05-06OR23177 under 
which the Jefferson Science Associates, LLC, operates the Thomas Jefferson National 
Accelerator Facility, the National Science Foundation, the French Atomic Energy 
Commission and National Center of Scientific Research, and the GDR n$^{\circ}$3034 
Physique du Nucl\'eon.

\end{theacknowledgments}

\bibliographystyle{aipproc}

\end{document}